\documentclass[aps,prd,twocolumn,superscriptaddress]{revtex4}
\usepackage{epsfig,epsf}
\usepackage{amsmath}
\usepackage{amsthm}
\usepackage{amsfonts}
\usepackage{amssymb}
\usepackage{dsfont}
\usepackage{multirow}
\usepackage{appendix}
\usepackage{slashed}
\usepackage[active]{srcltx}
\usepackage{psfrag}

\begin{document}

\title{Semileptonic $\Lambda_{b}\rightarrow
\Lambda_{c}{\ell}\bar\nu_{\ell}$ Transition in Full QCD}
\date{\today}
\author{K.~Azizi}
\affiliation{School of Physics, Institute for Research in
Fundamental Sciences (IPM), P.~O.~Box 19395-5531, Tehran, Iran}
\affiliation{Physics Department, Do\u gu\c s University,
Ac{\i}badem-Kad{\i}k\"oy, 34722 Istanbul, Turkey}
\author{J. Y. S\"{u}ng\"{u}}
\affiliation{Department of Physics, Kocaeli University, 41380
Izmit, Turkey}

\begin{abstract}

The tree-level $b\rightarrow
c{\ell}\bar\nu_{\ell}$ based hadronic transitions have been on the focus of much attention since recording significant deviations of the experimental data, on the ratios of the branching fractions in $ \tau $ and $ e-\mu $ channels of the semileptonic $ B \rightarrow D $ transition, from the SM predictions by BaBar Collaboration in 2012. It can be of great importance to look whether similar discrepancies take place in the semileptonic baryonic $\Lambda_{b}\rightarrow
\Lambda_{c}{\ell}\bar\nu_{\ell}$ decay channel or not. In this accordance we estimate the decay width as well as the ratios of the branching fractions in $ \tau $ and $ e-\mu $ channels of this baryonic transition by calculating the form factors, entering the amplitude of this transition as the main inputs, in the framework of QCD sum rules in full  theory.  We compare the obtained results with the predictions of other theoretical studies. Our results may be compared with the corresponding future experimental data to look for possible deviations of data from the SM predictions.

\end{abstract}

\maketitle

\section{Introduction}
One of the main goals of  the LHC, after the discovery of Higgs, is
to search for the new physics (NP) effects. This is done via two
ways: direct search at colliders and indirect search for the NP
effects in hadronic decay channels. Recently, there have been
recorded significant deviations from the SM predictions: the BaBar
measurements \cite{Lees:2012xj} on the ratios of the branching
fractions of the semileptonic $B\rightarrow D$ decay in $\tau$
channel to those of the $e$ or $\mu$ had shown to deviate at the
global level of $3.4\sigma$ from the SM predictions
\cite{Fajfer:2012vx,Fajfer:2012jt}.
One of the most important   results newly obtained at LHC is the sign of
the lepton flavor universality violation (LFUV) in the
semileptonic $B$ decays.
The LHC data \cite{Aaij:2014ora} on
\begin{equation}\label{MuonElectronRatio}
R_{k}=\frac{Br(B^+\rightarrow K^+\mu\mu)}{Br(B^+\rightarrow K^+ee)}=0.745^{+0.090}_{-0.074} (stat)\pm
0.036(syst)
\end{equation}
lies $2.6\sigma$ below the SM prediction in the window
$q^2\in[1,6]~\mathrm{GeV^2}$. Similar indications have been newly
reported in semileptonic decay, $B\rightarrow K$
\cite{Aaij:2017vbb}. Hence, the semileptonic $B$ decays seem to be
good probe to search for the new physics beyond the SM. In
principle, similar behaviors and deviations from the SM predictions can occur in other b-hadron decays. In \cite{Azizi:2016dcj}, it was shown that some experimental data on the differential branching ratio as well as lepton forward-backward asymmetry in $\Lambda_{b}\rightarrow \Lambda \mu^{+}\mu^{-}$ channel can not be described by the SM 
 predictions provided by the lattice QCD and QCD sum rules. Although there previously were predictions of different models in heavy quark effective theory limit,  the form factors of $\Lambda_{b}\rightarrow \Lambda \ell^{+}\ell^{-}$ 
 were firstly calculated in 2010 in full theory including all twelve form factors in Ref. \cite{Aliev:2010uy}. The obtained results on the order of branching fractions at different lepton channels had shown that these channels were accessible at hadron colliders. One year later, the CDF collaboration observed this baryonic flavor-changing neutral current decay in $ \mu $ channel \cite{Aaltonen:2011qs}. In 2015 the LHCb collaboration measured the 
 differential branching ratio and made angular analysis of the same decay mode \cite{Aaij:2015xza}. In 2016, the lattice predictions became available, where the form factors,
differential branching fraction, and angular observables  with relativistic b quarks associated to this channel were calculated \cite{Detmold:2016pkz}.
Considering the new experimental developments on the spectroscopy and decays properties of heavy hadrons, it seems that the b-baryon decays, especially the $\Lambda_b$ baryon decay modes become important not only for exact determinations of different SM parameters  but as essential sources of the physics BSM:  very recently  the LHCb Collaboration has found evidence for CP violation in $ \Lambda_{b} $ to $ p\pi^{-} \pi^{+} \pi^{-}$ decays with a
statistical significance corresponding to 3.3 standard deviations including systematic uncertainties. This represents the first
evidence for CP violation in the baryon sector \cite{Aaij:2016cla}.

 Many parameters related to different decay channels of the  $\Lambda_b$ state have been previously  studied using different approaches such as relativistic quark model, soft-collinear effective theory, heavy quark effective theory, covariant quark model, zero recoil sum rule, lattice QCD and QCD sum rules (see for instance Refs.  \cite{Feldmann:2011xf,MarquesdeCarvalho:1999bqs,Aliev:2002tr,Aliev:2010uy,Azizi:2009wn,Detmold:2015aaa,Mannel:2015osa,Faustov:2016pal,Gutsche:2015mxa,Dutta:2015ueb,gikls,prc,wkl,Datta:2017aue,Dai:1996xv} and references therein). 
The tree-level $b\rightarrow c{\ell}\bar\nu_{\ell}$ based semileptonic $\Lambda_{b}\rightarrow
\Lambda_{c}{\ell}\bar\nu_{\ell}$ transition is one of the prominent decay channels of the $ \Lambda_{b} $ baryon. This channel has  been investigated using different quark models and lattice QCD  \cite{Detmold:2015aaa,Mannel:2015osa,Faustov:2016pal,Gutsche:2015mxa,Dutta:2015ueb,gikls,prc,wkl,Datta:2017aue}. We analyze this decay in $ e $, $ \mu $ and $ \tau $ channels. In particular we calculate all six form factors entering the the matric elements of the effective Hamiltonian sandwiched between the initial and final baryonic states in full QCD without making the heavy quark effective theory approximation. We calculate the decay width and branching ratios in all lepton channels and compare the results with the predictions of other theoretical approaches as well as existing experimental data. We compute the  ratio of the branching
fractions  in $\tau$
channel to those of the $e$ or $\mu$ associated to this transition, as well.

This paper is organized as follows. In  section II, we calculate the six form factors defining the  $\Lambda_{b}\rightarrow
\Lambda_{c}{\ell}\bar\nu_{\ell}$ transition using the technique of QCD sum rules \cite{Shifman:1978by}. In section III we numerically analyze the form factors and find  their $ q^{2} $-dependent fit functions. 
Section IV is devoted to calculations of different physical observables related to the decays under consideration and comparison of the results obtained with the predictions of other theoretical studies as well as existing experimental data. Section V is reserved for our concluding remarks and,  finally, we move some analytic expressions for the spectral densities used in the calculations to the Appendix.

\section{Transition Form Factors}

The $\Lambda_{b}\rightarrow \Lambda_{c}{\ell}\bar\nu_{\ell}$ decay
channel proceeds via  $b\rightarrow
c{\ell}\bar\nu_{\ell}$ at quark level.
The low-energy effective Hamiltonian describing this transition can be written as
\begin{eqnarray}\label{Heff}
{\cal H}_{eff} =
\frac{G_F}{\sqrt2} V_{cb} ~\bar c \gamma_\mu(1-\gamma_5)~b~\bar l
~\gamma^\mu(1-\gamma_5) \nu,
\end{eqnarray}
where $G_F$ is the Fermi coupling constant and $V_{cb}$ is one of
the elements of the CKM matrix. The amplitude of this channel is found by sandwiching this effective Hamiltonian between the initial and final baryonic state,
\begin{eqnarray}\label{amp}
M=\langle \Lambda_{c}\vert{\cal H}_{eff}\vert \Lambda_{b}\rangle,
\end{eqnarray}
where the point-like particles immediately go out of the matrix element and remaining parts are parameterized in terms of six form factors $ F_1(q^2),~F_2(q^2),~ F_3(q^2)$ and $ G_1(q^2),~G_2(q^2),~ G_3(q^2)$ in full QCD:

\begin{eqnarray}\label{Cur.with FormFac.}
&&\langle \Lambda_c(p',s')|V^{\mu}|\Lambda_b (p,s)\rangle = \bar
u_{\Lambda_c}(p',s') \notag \\
&&\times
\Big[F_1(q^2)\gamma^{\mu}+F_2(q^2)\frac{p^{\mu}}{M_{\Lambda_b}}
+F_3(q^2)\frac{p'^{\mu}}{M_{\Lambda_c}}\Big] u_{\Lambda_b}(p,s), \notag \\
&&\langle \Lambda_c(p',s')|A^{\mu}|\Lambda_b (p,s)\rangle = \bar
u_{\Lambda_c}(p',s')     \notag \\
&&\times\Big[G_1(q^2)\gamma^{\mu}+G_2(q^2)\frac{p^{\mu}}{M_{\Lambda_b}}+G_3(q^2)\frac{p'^{\mu}}{M_{\Lambda_c}}\Big]
\gamma_5 u_{\Lambda_b}(p,s). \notag \\
\end{eqnarray}

In above equation,  $V^\mu=\overline{c}\gamma_\mu b$ and $A^\mu= \overline{c}\gamma_\mu \gamma_5 b$ are the vector and axial vector parts of the transition current, $q=p-p'$ is the
momentum transferred to the leptons; and  $u_{\Lambda_{c}}(p,s)$ and
$u_{\Lambda_{c}}(p',s')$ are Dirac spinors of the initial and
final baryonic states.

The main goal in the following is to calculate the six transition form factors in full QCD using the technique of the three-point sum rule as one of the powerful and applicable non-perturbative tools to hadron physics. As usual prescriptions, the starting point  is to consider an appropriate correlation function of interpolating and transition currents in a time ordered manner.
The sum rules for transition form factors are found by equating the phenomenological or physical
representation of this three point function to the theoretical
or QCD side of the same function which is obtained using the operator product expansion (OPE). The three-point correlation function for our aim is: 
\begin{eqnarray}\label{CorFunc}
\Pi_{\mu}(p,p^{\prime},q)&=&i^2\int d^{4}x e^{-ip\cdot
x}\int d^{4}y e^{ip'\cdot y} \notag \\
&\times& \langle 0|{\cal T}|{\cal J}^{\Lambda_c}(y){\cal
J}_{\mu}^{tr,V(A)}(0) {\cal J}^{\dag \Lambda_b}(x)|0\rangle,
\end{eqnarray}
where $\cal T$ is the time-ordereing operator and $ {\cal J}^{\Lambda_{Q}}(x) $ with $ Q $ being $ b $ or $ c $ quark is  the interpolating current for the $ \Lambda_{b} $ and  $ \Lambda_{c} $ baryons. It is given in its more general form as:
\begin{widetext}
\begin{eqnarray} \label{Current}
{\cal J}^{\Lambda_{Q}}(x)&=&\frac{1}{\sqrt{6}}~\epsilon_{abc}
\Bigg\{2\Big[\Big(q_{1}^{aT}(x)Cq_{2}^{b}(x)\Big)\gamma_{5}Q^{c}(x)
+\beta\Big(q_{1}^{aT}(x)C\gamma_{5}q_{2}^{b}(x)\Big)Q^{c}(x)\Big]
+\Big(q_{1}^{aT}(x)CQ^{b}(x)\Big)\gamma_{5}q_{2}^{c}(x)    \notag \\
&+&\beta\Big(q_{1}^{aT}(x)C\gamma_{5}Q^{b}(x)\Big)q_{2}^{c}(x)
+\Big(Q^{aT}(x)Cq_{2}^{b}(x)\Big)\gamma_{5}q_{1}^{c}(x)
+\beta\Big(Q^{aT}(x)C\gamma_{5}q_{2}^{b}(x)\Big)q_{1}^{c}(x)\Bigg\}~,
\end{eqnarray}
where $a,~b$ and $c$ are color indices, $C$ is the charge
conjugation operator, $q_{1}$ and $q_{2}$ are  $u$ and $d$
quark fields, respectively. The $\beta$ is a general mixing parameter with $\beta=-1$
being corresponding to  Ioffe current. The physical or phenomenological side is found by inserting complete sets of the initial and final baryonic states with the same quantum numbers as the interpolating currents into the correlation function. By performing integrals over four-$ x $ and -$ y $ we end up with
\begin{eqnarray} \label{PhysSide}
\Pi_{\mu}^{Phys.}(p,p',q)=\frac{\langle 0 \mid {\cal
J}^{\Lambda_c} (0)\mid \Lambda_c(p') \rangle \langle \Lambda_{c} (p')\mid
{\cal J}_{\mu}^{tr,V(A)}(0)\mid \Lambda_b(p) \rangle \langle \Lambda_{b}(p)
\mid {\cal J}^{\dag \Lambda_b}(0)\mid
0\rangle}{(p'^2-m_{\Lambda_c}^2)(p^2-m_{\Lambda_b}^2)}+\cdots~,
\end{eqnarray}
\end{widetext}
where $\cdots$ stands for the contributions of the higher states and
 continuum. Besides the transition matrix elements we need to define the following matrix elements in terms of the  residues of the initial
  and final states: 
\begin{eqnarray}\label{MatrixElements}
\langle 0|{\cal J}^{\Lambda_c}(0)|\Lambda_c(p')\rangle =
\lambda_{\Lambda_c} u_{\Lambda_c}(p',s'), \notag \\
\langle\Lambda_b(p)|\bar {\cal J}^{\Lambda_b}(0)| 0 \rangle =
\lambda^{+}_{\Lambda_b}\bar u_{\Lambda_b}(p,s).
\end{eqnarray}
The final step is to put  all the matrix elements defined above into Eq. (\ref{PhysSide}) and use the summation over Dirac spinors
\begin{eqnarray}\label{Spinors}
u_{\Lambda_c} (p',s')~\bar{u}_{\Lambda_c}
(p',s')&=&\slashed{p}~'+M_{\Lambda_c},\notag \\
u_{\Lambda_b}(p,s)~\bar{u}_{\Lambda_b}(p,s)&=&\slashed
p+M_{\Lambda_b}.
\end{eqnarray}
As a result we find the following representation for the final form of the physical side in terms of the structures used in the calculations in Borel transformed form that has been applied to suppress the contributions of the higher states and continuum: 
\begin{eqnarray}\label{Physical Side structures}
&&\mathbf{\widehat{B}}\Pi_{\mu}^{\mathrm{Phys.}}(p,p',q)=\Bigg[m_{\Lambda_b}
F_{1} \slashed{p}' \gamma_{\mu}+\frac{1}{m_{\Lambda_b}}F_{2}
p_{\mu}
\slashed {p}' \slashed {p} \notag \\
&&+\frac{1}{m_{\Lambda_c}}F_{3} p'_{\mu} \slashed {p}' \slashed
{p}+ m_{\Lambda_b} m_{\Lambda_c} G_{1}
\gamma_{\mu}\gamma_{5} \notag \\
&&-\frac{1}{m_{\Lambda_b}}G_{2} p_{\mu} \slashed {p}' \slashed
{p} ~\gamma_{5}-\frac{1}{m_{\Lambda_b}} G_{3} p'_{\mu}
\slashed {p}~'\slashed{p}~\gamma_{5}+...\Bigg] \notag \\
&& \times
\lambda_{\Lambda_b}\lambda_{\Lambda_c}~~e^{-\frac{m_{\Lambda_b}^2}{M^2}}
~~e^{-\frac{m_{\Lambda_c}^2}{M'^{2}}} ,
\end{eqnarray}
where $ M^2 $ and $M'^{2}$ are Borel parameters that should be fixed later and we kept only
the structures that will be used in further analyses.

To find the correlation function  in terms of quark-gluon
degrees of freedom on QCD side, i.e. by utilizing the light and heavy
propagators, we use the interpolating current given by Eq.\
(\ref{Current}) in Eq.\ (\ref{CorFunc}), and contract the
related quark fields. After some manipulations including the contraction of the quark fields, we find the QCD  side in terms of the heavy and light quarks' propagators in coordinate space. Thus, for the light quark we use
\begin{eqnarray}\label{LightProp}
&&S_{q}^{ab}(x)=i\delta _{ab}\frac{\slashed x}{2\pi ^{2}x^{4}}-\delta _{ab}%
\frac{m_{q}}{4\pi ^{2}x^{2}}-\delta _{ab}\frac{\langle
\overline{q}q\rangle
}{12}  \notag \\
&&+i\delta _{ab}\frac{\slashed xm_{q}\langle \overline{q}q\rangle }{48}%
-\delta _{ab}\frac{x^{2}}{192}\langle \overline{q}g_{}\sigma
Gq\rangle
+i\delta _{ab}\frac{x^{2}\slashed xm_{q}}{1152}\langle \overline{q}%
g_{}\sigma Gq\rangle  \notag \\
&&-i\frac{g_{}G_{ab}^{\alpha \beta }}{32\pi ^{2}x^{2}}\left[ \slashed x{%
\sigma _{\alpha \beta }+\sigma _{\alpha \beta }}\slashed x\right]
-i\delta _{ab}\frac{x^{2}\slashed xg_{}^{2}\langle
\overline{q}q\rangle ^{2}}{7776} \notag \\
&&-\delta _{ab}\frac{x^{4}\langle \overline{q}q\rangle \langle
g_{}^{2}G^{2}\rangle }{27648}+\ldots,
\end{eqnarray}
and the heavy quark propagator is given as \cite{Reinders};
\begin{eqnarray}\label{HeavyProp}
&&S_{Q}^{ab}(x)=i\int \frac{d^{4}k}{(2\pi )^{4}}e^{-ikx}\Bigg
\{\frac{\delta_{ab}\left( {\slashed k}+m_{Q}\right) }{k^{2}-m_{Q}^{2}}    \notag \\
&&-\frac{g_{}G_{ab}^{\alpha \beta }}{4}\frac{\sigma _{\alpha \beta }\left( {%
\slashed k}+m_{Q}\right) +\left( {\slashed k}+m_{Q}\right) \sigma
_{\alpha\beta }}{(k^{2}-m_{Q}^{2})^{2}}   \notag \\
&&+\frac{g_{}^{2}G^{2}}{12}\delta _{ab}m_{Q}\frac{k^{2}+m_{Q}{\slashed k}}{%
(k^{2}-m_{Q}^{2})^{4}}+\ldots %
\Bigg \},
\end{eqnarray}%
where we used the following notations
\begin{eqnarray}\label{GluonField}
&&G_{ab}^{\alpha \beta }=G_{A}^{\alpha \beta
}t_{ab}^{A},\,\,~~G^{2}=G_{\alpha \beta }^{A}G_{\alpha \beta
}^{A},
\end{eqnarray}
with $a,\,b=1,2,3$ being the color and $A,B,C=1,\,2\,\ldots 8$
being the flavor indices. In Eq.\ (\ref{GluonField}) $t^{A}=\lambda ^{A}/2$,
$\lambda ^{A}$ are the Gell-Mann matrices and the gluon field
strength tensor $G_{\alpha \beta}^{A}\equiv G_{\alpha \beta
}^{A}(0)$ is fixed at $x=0$.

By replacing the heavy and light quark propagators we apply  the following Fourier
transformation:
\begin{eqnarray}\label{intyx}
\frac{1}{[(y-x)^2]^n}&=&\int\frac{d^Dt}{(2\pi)^D}e^{-it\cdot(y-x)}~i~(-1)^{n+1}~2^{D-2n}~\pi^{D/2} \notag\\
&&\times \frac{\Gamma(D/2-n)}{\Gamma(n)}\Big(-\frac{1}{t^2}\Big)^{D/2-n}.
\end{eqnarray}
Then, the four-dimensional $x$ and $y$ integrals are performed in the
sequel of the replacements $x_{\mu}\rightarrow
i\frac{\partial}{\partial p_{\mu}}$ and $y_{\mu}\rightarrow
-i\frac{\partial}{\partial p'_{\mu}}$. This procedure brings two four-dimensional Dirac delta functions which are used to perform the
four-integrals over $k$ and $k^{\prime}$ coming from the heavy $b$ and $c$ quarks propagators. Then the Feynman
parametrization and
\begin{eqnarray}\label{Int}
\int d^4t\frac{(t^2)^{\beta}}{(t^2+L)^{\alpha}}=\frac{i \pi^2
(-1)^{\beta-\alpha}\Gamma(\beta+2)\Gamma(\alpha-\beta-2)}{\Gamma(2)
\Gamma(\alpha)[-L]^{\alpha-\beta-2}},\quad
\end{eqnarray}
are used to carry out the remaining four-integral over $t$.
The function $\Pi_{\mu}^{\mathrm{QCD}}(p,p',q)$ includes twenty-four
different structures that not all of them are written here:
\begin{eqnarray}\label{Structures}
\Pi_{\mu}^{\mathrm{QCD}}(p,p',q)&=&\Pi^{\mathrm{QCD}}_{\slashed{p}' \gamma_{\mu}}(p^{2},p'^{2},q^{2})\slashed{p}' \gamma_{\mu}\nonumber\\&+&
\Pi^{\mathrm{QCD}}_{p_{\mu} \slashed {p}~'
\slashed {p}}(p^{2},p'^{2},q^{2})p_{\mu} \slashed {p}~'
\slashed {p}+...., \nonumber\\
\end{eqnarray}
where, the invariant functions $\Pi
^{\mathrm{QCD}}_i(p^{2},p'^{2},q^{2})$,  with $ i $ representing different structures, are
represented in terms of a double dispersion integral as
\begin{eqnarray}\label{PiQCD}
\Pi^{\mathrm{QCD}}_i(p^{2},p'^{2},q^{2})&=&\int_{s_{min}}^{\infty}ds
\int_{s'_{min}}^{\infty}ds'~\frac{\rho
^{\mathrm{QCD}}_i(s,s',q^{2})}{(s-p^{2})(s'-p'^{2})} \notag\\
&+&\mathrm{subtracted~terms},
\end{eqnarray}
where $s_{min}=(m_u+m_d+m_b)^{2}$, $s'_{min}=(m_u+m_d+m_c)^{2}$ and $\rho_i^{\mathrm{QCD}}(s,s',q^{2})$ are
the  spectral densities corresponding to different structures. These spectral densities that are obtained by taking the imaginary parts of the $ \Pi^{\mathrm{QCD}}_i(p^{2},p'^{2},q^{2}) $ functions according to the standard prescriptions of the method used,  include  two
different parts and can be classified as
\begin{equation} \label{Rho}
\rho^{\mathrm{QCD}}_i(s,s',q^{2})=\rho_i ^{Pert.}(s,s',q^{2})+\sum_{n=3}^{5}\rho_{i}^{n}(s,s',q^{2}),
\end{equation}
where by $\rho_{i}^{n}(s,s',q^{2})$ we denote the nonperturbative contributions
to $\rho^{\mathrm{QCD}}_i(s,s',q^{2})$: $ n=3$, $ 4 $ and $ 5 $ stand for the quark, gluon  and mixed condensates, respectively.
 Due to the lengthy
expressions of the spectral densities, we present only the explicit forms of the spectral
densities $\rho^{Pert.}_{\slashed{p}' \gamma_{\mu}}(s,s^{\prime},q^2)$ and
$\rho^{n}_{\slashed{p}' \gamma_{\mu}}(s,s^{\prime},q^2)$ corresponding to the Dirac
structure $\slashed{p}' \gamma_{\mu}$  in Appendix.

After applying the double Borel transformation on the variables $p^{2}$ and $p'^{2}$  in QCD side  and subtracting
the contribution of the higher resonances and continuum states
supported by the quark-hadron duality assumption and matching the coefficients of different structures from the physical and QCD sides of the correlation function, we find the required  sum rules  for the form
factors that will be used in numerical calculations.

\section{Numerical results for form factors}

In this section, we shall give our numerical results for the form
factors and find their fit functions in terms of $ q ^2$. In our calculations,  we set
$m_u$ and $m_d$ equal to zero. Other input
parameters used in our evaluation are collected in
Table I.
\begin{table}[h!]
\label{inputPar}
\caption{Input parameters used in 
calculations.}
\begin{tabular}{|c|c|}
\hline \hline
Parameters                                             &  Values  \\
\hline \hline
$ m_c$                                                 & $(1.28\pm0.03)~ \mathrm{GeV}$ \cite{Patrignani:2016xqp}\\
$ m_b$                                                 & $(4.18^{0.04}_{2.29})~ \mathrm{GeV}$ \cite{Patrignani:2016xqp}\\
$ m_e $                                                & $ 0.00051~~\mathrm{GeV}$ \cite{Patrignani:2016xqp}\\
$ m_\mu $                                              & $ 0.1056~~\mathrm{GeV}$ \cite{Patrignani:2016xqp}\\
$ m_\tau $                                             & $ 1.776~~\mathrm{GeV}$ \cite{Patrignani:2016xqp}\\
$ M_{\Lambda_b}$                                       & $ (5619.51\pm0.23) \mathrm{MeV}$   \cite{Patrignani:2016xqp}\\
$ M_{\Lambda_c} $                                      & $ (2286.46\pm0.14)\mathrm{GeV}$  \cite{Patrignani:2016xqp} \\
$ G_{F} $                                              & $ 1.17\times 10^{-5} \mathrm{GeV^{-2}}$ \cite{Patrignani:2016xqp}\\
$ V_{cb} $                                             & $ (39\pm1.1)\times 10^{-3}$  \cite{Patrignani:2016xqp}\\
$ m^2_0 $                                              & $ (0.8\pm0.2)\mathrm{GeV^2}$ \cite{Dosch:1988vv,Belyaev:1982cd}\\
$\tau_{\Lambda_b} $                                    & $ 1.47\times 10^{-12}$ \cite{Patrignani:2016xqp}\\
$\langle u\bar{u}\rangle$ $ = $ $\langle d\bar{d}\rangle$ & $(0.24\pm0.01)^3 \mathrm{GeV^3}$ \cite{Ioffe:2005ym} \\
$\langle0|\frac{1}{\pi}\alpha_s G^2|0\rangle$          &$ (0.012\pm0.004)\mathrm{GeV^4}$ \cite{Ioffe:2005ym}\\
\hline\hline
\end{tabular}
\end{table}
The sum rules for form factors  contain extra four auxiliary parameters: the Borel parameters $M^2$
and $M'^{2}$ as well as the continuum thresholds $s_0$ and
$s'_0$. According to the standard prescriptions of the method, the results of 
form factors should be practically independent of these parameters. Hence
their working regions are settled such that the results of 
form factors depend possibly weakly on these parameters.

 The continuum thresholds $s_{0}$  and
$s'_0$ are  not
entirely arbitrary parameters and they are in correlation with the energy of the
first excited states in the initial and final channels. We choose them  the intervals
$$(m_{\Lambda_b}+0.1)^2~ \mathrm{GeV^2} \leq s_{0} \leq
(m_{\Lambda_b}+0.5)^2~ \mathrm{GeV^2},$$ and $$
(m_{\Lambda_c}+0.1)^2~\mathrm{GeV^2}\leq s'_{0} \leq
(m_{\Lambda_c}+0.5)^2~ \mathrm{GeV^2}.$$

The working regions for the Borel mass parameters are determined with
the requirements that not only the higher state and continuum
contributions are suppressed but also the contributions of the
higher order operators are small, i.e. the sum rules are
convergent. Accordingly, the working regions for the Borel
parameters are found to be $$ 6~\mathrm{GeV^2}\leq M^2 \leq
10~\mathrm{GeV^2}, $$ and $$4~\mathrm{GeV^2} \leq M'^2 \leq
6~\mathrm{GeV^2}.$$

The aforesaid intervals for the Borel and threshold parameters
prompts  the below window to the parameter $\beta$:
$$-0.5\leq x \leq +0.5,$$ where we utilize $x = cos\theta$ with
$\theta = tan^{-1}\beta$  to examine the full region 
i.e.$-\infty$ to $\infty$ for $\beta$ by changing $ x $ in the interval $[-1, 1]$.
Note that the Ioffe current with $x = -0.71$ stays out of the
trustworthy region in this evaluation. 

Having determined the working regions for the auxiliary parameters
we proceed to find the behaviors of the form factors in terms of
$q^2$. Our analysis shows that the form factors are well fitted to
the function
\begin{equation} \label{fitff}
{\cal F}(q^2)=\frac{{\cal
F}(0)}{\displaystyle\left(1-\xi_1\frac{q^2}{M^2_{\Lambda_Q}}+\xi_2
\frac{q^4}{M_{\Lambda_Q}^4}+\xi_3\frac{q^6}{M_{\Lambda_Q}^6}+\xi_4\frac{q^8}{M_{\Lambda_Q}^8}\right)},
\end{equation}
where the values of the parameters, ${\cal F}(0)$, $\xi_1$, $\xi_2$,
$\xi_3$ and $\xi_4$ obtained using $M^2=8~\mbox{GeV$^2$}$ and
$M'^2=5~\mbox{GeV$^2$}$ for $\Lambda_b\rightarrow \Lambda_c \ell
\bar{\nu}$ transition are presented in the Table~
II.
\begin{widetext}

\begin{table}
\label{Fitresults:j}
\caption{Parameters of the fit function for different form factors corresponding to
$\Lambda_b\to\Lambda_c$ transition.}
\begin{ruledtabular}
\begin{tabular}{|c|c|c|c|c|c|c|}
                  & $F_1(q^2)$      & $F_2(q^2)$        & $F_3(q^2)$       & $G_1(q^2)$       & $G_2(q^2)$        & $G_3(q^2)$       \\
\hline
${\cal F}(q^2=0)$ & $1.220\pm0.293$ & $-0.256\pm0.061$  & $-0.421\pm0.101$ & $0.751\pm0.180$  & $-0.156\pm0.037$  & $0.320\pm0.077$  \\
$\xi_1$           & $1.03$          & $2.17$            & $2.18$           & $1.41$           & $1.46$            & 2.36              \\
$\xi_2$           & $-4.60$         & $-8.63$           & $-1.02$          & $-3.30$          & $-6.50$           & $-2.90$           \\
$\xi_3$           & $28$            & $51.40$           & $18.12$          & $21.90$          & $41.20$           & $28.20$           \\
$\xi_4$           & $-53$           & $-85.2$           & $-32$            & $-40.10$         & $-74.82$          & $-45.2$           \\
\end{tabular}
\end{ruledtabular}
\end{table}
\begin{figure}[h!] \label{Fig:FQsq}
\includegraphics[totalheight=5cm,width=5.8cm]{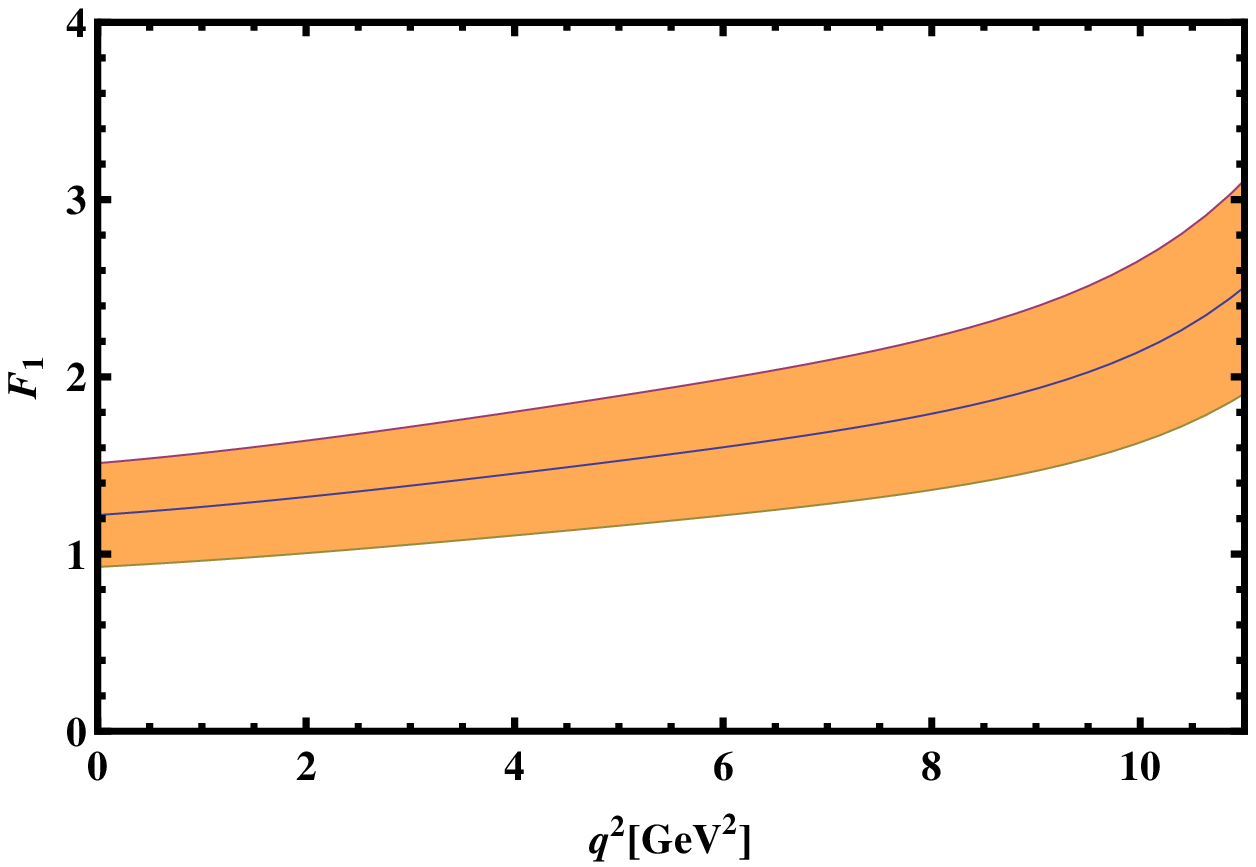}
\includegraphics[totalheight=5cm,width=5.8cm]{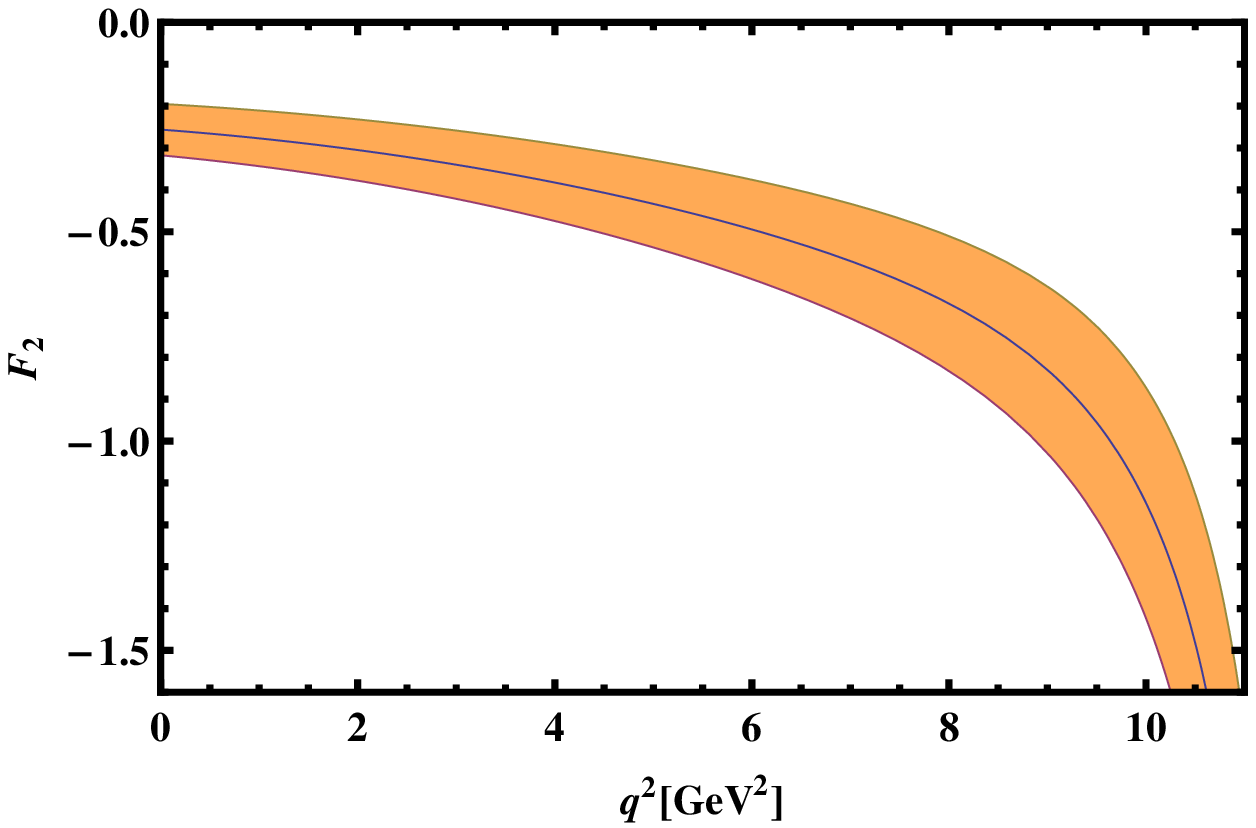}
\includegraphics[totalheight=5cm,width=5.8cm]{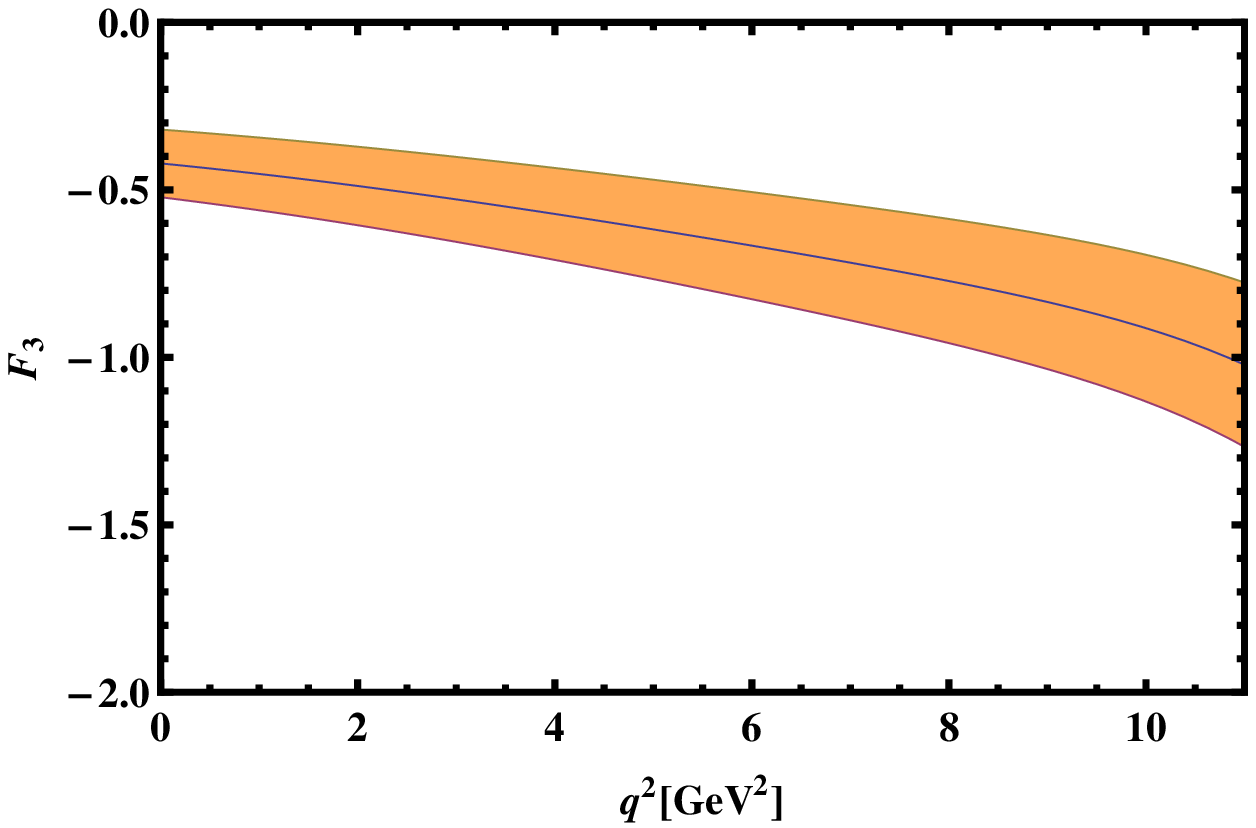}
\caption{$F_1, F_2$ and $F_3$ form factors as a function of
 $q^2$ at average values of auxiliary parameters.}
\end{figure}
\begin{figure}[h!] \label{Fig:GQsq}
\includegraphics[totalheight=5cm,width=5.8cm]{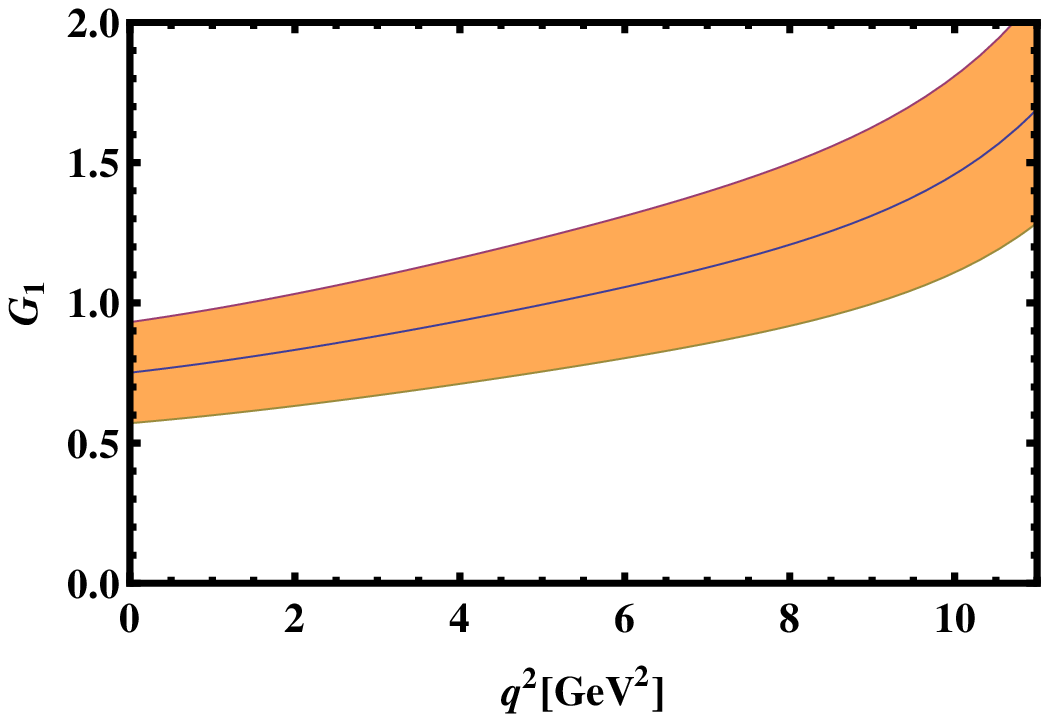}
\includegraphics[totalheight=5cm,width=5.8cm]{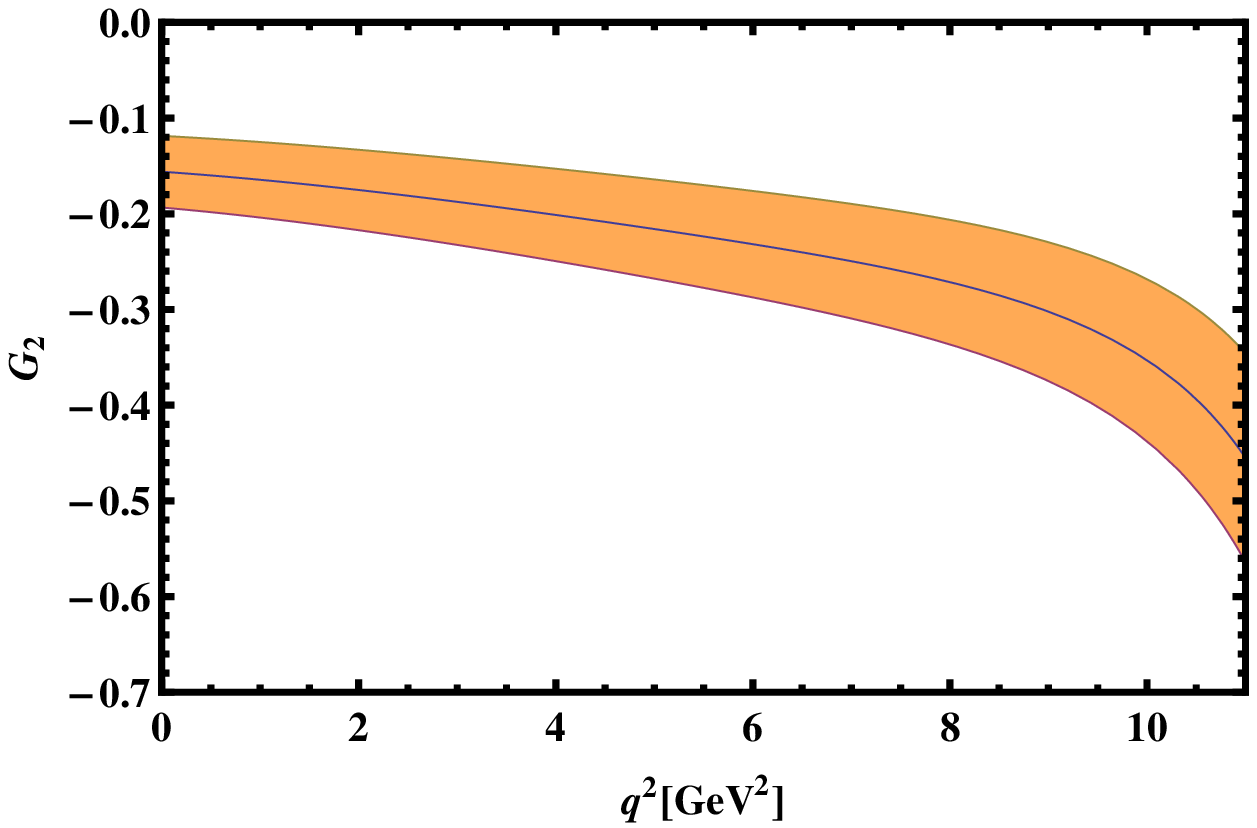}
\includegraphics[totalheight=5cm,width=5.8cm]{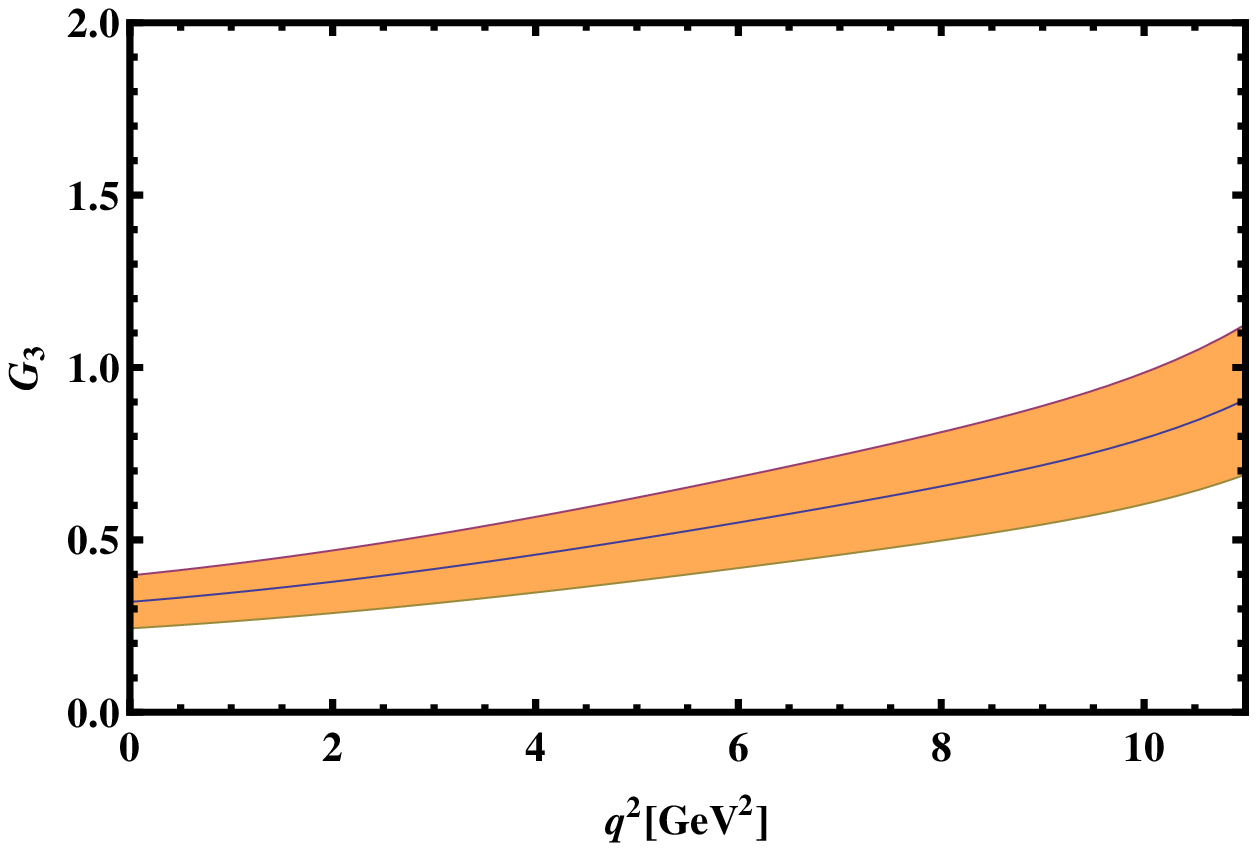}
\caption{$G_1, G_2$ and $G_3$ form factors as a function of
 $q^2$ at average values of auxiliary parameters.}
\end{figure}

\end{widetext}
Figures 1 and 2 show the dependence of the form factors $ F_i $ and $ G_i $ on $q^2$ in its allowed region, $ m_l^2\leq q^2 \leq
(m_{\Lambda_b}-m_{\Lambda_c})^2$ and at average values of auxiliary parameters.  As is seen we encounter the uncertainties of the form factors in these figures. The solid lines show  the average behavior of the form factors. From these figures we see that the form factors demonstrate a good behavior and gradually increase with increasing the transferred momentum squared. The fit functions of form factors will be used as the main input parameters to evaluate different physical observables in next section.



\section{Decay Width and Branching Ratio of $\Lambda_{b}\rightarrow\Lambda_{c}{\ell}\bar{\nu_{\ell}}$}

In this section we would like to evaluate  the decay widths and branching fractions of the
semileptonic $\Lambda_b\rightarrow  \Lambda_b\bar{\ell}{\nu}$ transitions in all lepton channels.
To this end we use the following formula:
\begin{equation}\label{eq:dgamma}
\frac{d\Gamma(\Lambda_b\to\Lambda_c\ell\bar\nu_\ell)}{dq^2}=\frac{G_F^2}{(2\pi)^3}
|V_{cb}|^2\frac{\lambda^{1/2}(q^2-m_\ell^2)^2}{48M_{\Lambda_b}^3q^2}{\cal
H}_{tot}(q^2),
\end{equation}
where,
\begin{equation}\label{eq:hh}
{\cal H}_{tot}(q^2)=[{\cal H}_U(q^2)+{\cal H}_L(q^2)]
\left(1+\frac{m_\ell^2}{2q^2}\right)+\frac{3m_\ell^2}{2q^2}{\cal
H}_S(q^2) .
\end{equation}
One can find the explicit expressions of the functions $ {\cal H}_{,U,L,S} $  in \cite{kk,Faustov:2016pal,gikls} in terms of different helicity amplitudes expressed 
as functions of the form factors $ F_i $ and $ G_i $.
The numerical values for the decay widths and branching ratios at different channels are shown in Table \ref{compdecaywidth}. In this table, we also present the predictions of other theoretical methods (in some cases we have changed the original unit to $ GeV $) as well as the existing experimental data. From this table we see that the order of magnitude for the widths and branching fractions from different theoretical predictions are the same, though they show considerable differences in values in some cases. Our result on the branching ratio in $ e, \mu $ channel is in nice agreement with average experimental value presented in PDG \cite{Patrignani:2016xqp}. Our predictions at $ \tau $ channel can be checked by future experiments.

As a final task we would like to report the ratio of branching fraction in $ \tau $ channel to that of the $ e, \mu $:
\begin{equation}
R=\frac{Br[\Lambda_b\rightarrow \Lambda_c\tau
\overline{\nu}_\tau]}{Br[\Lambda_b\rightarrow \Lambda_c
(e,\mu)\overline{\nu}_{(e,\mu)}]}=0.31\pm 0.11,
\end{equation}
which may  also be checked by future experiments.
\begin{widetext}

\begin{table}[bth!]
\caption{Decay width (in $\mathrm{GeV}  $) and branching ratio of the semileptonic $\Lambda_b\to\Lambda_c \ell {\overline{\nu}}_\ell$ transition.} \label{compdecaywidth} 
\begin{ruledtabular}
\begin{tabular}{ccccccccc}
Parameter&Present Work&\cite{Faustov:2016pal}&\cite{Gutsche:2015mxa,gikls}& \cite{prc}&\cite{Dutta:2015ueb}&\cite{wkl}&\cite{Detmold:2015aaa}&Exp.\cite{Patrignani:2016xqp}\\
\hline
$\Lambda_b\to\Lambda_c (e,\mu) {\overline{\nu}}_{(e,\mu)}$\\
$\Gamma\times10^{14}$      &  $ 2.32 \pm0.64$    & 2.91   &    & 3.03  &      &           & 2.23  &                                   \\
$Br$ (\%)                                   & $ 6.04\pm 1.70 $     & 6.48   & 6.9   &       &4.83  &6.3        &           &$6.2^{+1.4}_{-1.3}$               \\
\hline
$\Lambda_b\to \Lambda_c \tau \overline{\nu}_{\tau}$\\
$\Gamma \times10^{15}$     &$ 7.35\pm2.06 $  & 9.15   &     & 1.25 &      &           & 7.34   &                                        \\
$Br$ (\%)                                   &$ 1.87\pm 0.52 $  & 2.03   &2.0  &      &  1.63    &           &                                        \\
\end{tabular}
\end{ruledtabular}
\end{table}

\end{widetext}

\section{Conclusion}
The recent serious deviations of the experimental data from the
theoretical productions made in the context of SM on the ratios of the
branching fractions of the mesonic $B\rightarrow D^{(*)}$ decays in $\tau$
channel to that of the $(e,\mu)$ have put this subject in the
focus of the much attention. While direct searches end up with
null results in the search of NP effects at different colliders,
these can be considered as significant indications for the NP
effects beyond the SM. The corresponding $b\rightarrow
 c\overline{\nu}_\ell$ based transition at baryon sector that is possible
to study in future experiments is the semileptonic
$\Lambda_b\rightarrow\Lambda_c\ell\overline{\nu}_\ell$ transition.
we shall look at different experiments whether similar deviations is the case in this
transition or not?
In this connection we studied this transition at all lepton
channels by calculating the responsible form factors in full QCD.
we used the fit functions of the form factors to estimate the
corresponding decay rates and branching fractions. We found the
ratio $R=\frac{Br[\Lambda_b\rightarrow \Lambda_c\tau
\overline{\nu}_\tau]}{Br[\Lambda_b\rightarrow \Lambda_c
(e,\mu)\overline{\nu}_{(e,\mu)}]}=0.31\pm 0.11,$ which may be checked in future
experiments. If we observe serious deviations of data on this ratio  from the 
 SM predictions, like those of the mesonic channels, this will increase our desire to
indirectly search for new physics effects in heavy hadronic decay
channels.

\appendix*

\section{The spectral densities}

\renewcommand{\theequation}{\Alph{section}.\arabic{equation}} \label{sec:App}
\begin{widetext}
In the following we present the explicit forms of spectral
densities corresponding to the  form factor  $F_1$:
\begin{eqnarray} \label{RhoPert}
\rho^{Pert.}_{\slashed{p}' \gamma_{\mu}}(s,s',q^2)&=&\int_{0}^{1}du \int_{0}^{1-u}dv
~~\Bigg[\beta^2\Bigg\{\frac{1}{1536 \pi^4 (u+v-1)^2} 
\Bigg[-16 m_{b}^3 (u-1) v (u+v)+16 m_b^2 m_c u v \notag \\
&\times&(u+v)+m_{b} \bigg(-7s'(u-1)^2 u^2+(u-1)  u\Big[8(s+s')-7(s+2s') u+q^2 (23u-8)\Big]v \notag \\
&+&\Big[-9s+8 (-3q^2+3 s+s') u+\Big(23 q^2-7(2s +s')\Big) u^2 \Big] v^2+s(9-7u) v^3-16 m_c^2 (u-1)u(u \notag \\
&+& v)\bigg)+m_c u \bigg(s(u+v-1)\Big[u+7uv+v(2 +7v)\Big]+u \Big[16m_c^2(u+v)+s'(u+v-1)(1+7u  \notag \\
&+&7v)\Big]-q^2u \Big[u+23uv+v (23 v-6)-1\Big]\bigg)\Bigg]\Bigg\}\notag \\ &+&\frac{1}{768\pi^4(u+v-1)^2}\Bigg\{-12m_b^3(u-1)v(u+v) + 12m_b^2 m_cuv(u+v)+m_b\Big[-7s'(u-1)^2u^2 \notag\\
&+&(u-1)u\Big(6(s+s')-7(s+2s')u+q^2(19u _ 6)\Big)v+\Big(-5s+6(-3q^2+3s+s')u+[19q^2 \notag\\
&-& 7(2s+s')]u^2\Big)v^2+s(5-7u) v^3-12m_c^2 (u-1)u(u+v)\Big]+m_c u \Big[q^2u \Big(u-19uv \notag\\
&+&(8-19v)v-1 \Big)+s(u+v-1)\Big(v(7v-2)+u(7v-1)\Big)+u\Big(12m_c^2(u+v)+s'(u+v  \notag\\
&-& 1)(7u+7v-1)\Big)\Big]\Bigg\}\Bigg]\Theta[L(s,s^{\prime},q^2)],
\end{eqnarray}
\begin{eqnarray} \label{Rho3}
\rho^3_{\slashed{p}' \gamma_{\mu}}(s,s',q^2)&=&\frac{1}{192\pi^2}\int_{0}^{1}du \int_{0}^{1-u}dv \Bigg\{\langle d\overline{d}\rangle
\Big(2\beta^2(4+3u)-\beta(4-12u)12u\Big)\notag \\
&+&\langle u\overline{u}\rangle
\Big(-3\beta^2(2+u)+4\beta(3u-1)+2\Big)\Bigg\}\Theta[L(s,s^{\prime},q^2)],
\end{eqnarray}
\begin{equation}\label{Rho4}
\rho^4_{\slashed{p}' \gamma_{\mu}}(s,s',q^2)=0,
\end{equation}
\begin{equation}\label{Rho5}
\rho^5_{\slashed{p}' \gamma_{\mu}}(s,s',q^2)=0,
\end{equation}
where,
\begin{eqnarray}\label{L}
L(s,s^{\prime},q^2)&=&- m_c^2 u + s^{\prime} u - s^{\prime}
u^2 - m_b^2 v + s v+ q^2 u v - s u v - s^{\prime} u v - s v^2
\end{eqnarray}
with $\Theta[...]$ being the unit-step function.

\section*{ACKNOWLEDGEMENTS}
We would like to thank H. Sundu for useful discussions. Work of K.A. was  partly financed by Dogu\c{s} University
through the project: BAP 2015-16-D1-B04.
\end{widetext}


\begin{thebibliography}{99}

\bibitem{Lees:2012xj}
  J.~P.~Lees {\it et al.} [BaBar Collaboration],
  Phys.\ Rev.\ Lett.\  {\bf 109}, 101802 (2012).

  \bibitem{Fajfer:2012vx}
  S.~Fajfer, J.~F.~Kamenik and I.~Nisandzic,
  Phys.\ Rev.\ D {\bf 85}, 094025 (2012).

  \bibitem{Fajfer:2012jt}
  S.~Fajfer, J.~F.~Kamenik, I.~Nisandzic and J.~Zupan,
  Phys.\ Rev.\ Lett.\  {\bf 109}, 161801 (2012).

\bibitem{Aaij:2014ora}
  R.~Aaij {\it et al.} [LHCb Collaboration],
  Phys.\ Rev.\ Lett.\  {\bf 113}, 151601 (2014).

\bibitem{Aaij:2017vbb}
  R.~Aaij {\it et al.} [LHCb Collaboration],
  JHEP {\bf 1708}, 055 (2017).
  
\bibitem{Azizi:2016dcj} 
  K.~Azizi, A.~T.~Olgun and Z.~Tavukoglu,
  Adv.\ High Energy Phys.\  {\bf 2017}, 7435876 (2017).
  
\bibitem{Aliev:2010uy}
T.~M.~Aliev, K.~Azizi and M.~Savci,
Phys.\ Rev.\ D {\bf 81}, 056006 (2010).

\bibitem{Aaltonen:2011qs}
T.~Aaltonen {\it et al.} [CDF Collaboration],
Phys.\ Rev.\ Lett.\  {\bf 107}, 201802 (2011).

\bibitem{Aaij:2015xza} 
  R.~Aaij {\it et al.} [LHCb Collaboration],
  JHEP {\bf 1506}, 115 (2015).

\bibitem{Detmold:2016pkz} 
  W.~Detmold and S.~Meinel,
  Phys.\ Rev.\ D {\bf 93}, no. 7, 074501 (2016).


\bibitem{Aaij:2016cla}
R.~Aaij {\it et al.} [LHCb Collaboration],
Nature Phys.\  {\bf 13}, 391 (2017).







\bibitem{Feldmann:2011xf}
T.~Feldmann and M.~W.~Y.~Yip,
Phys.\ Rev.\ D {\bf 85}, 014035 (2012)
Erratum: [Phys.\ Rev.\ D {\bf 86}, 079901 (2012)].

\bibitem{MarquesdeCarvalho:1999bqs}
R.~S.~Marques de Carvalho, F.~S.~Navarra, M.~Nielsen, E.~Ferreira and H.~G.~Dosch,
Phys.\ Rev.\ D {\bf 60}, 034009 (1999).

\bibitem{Aliev:2002tr}
T.~M.~Aliev, A.~Ozpineci and M.~Savci,
Phys.\ Rev.\ D {\bf 67}, 035007 (2003).



\bibitem{Azizi:2009wn}
K.~Azizi, M.~Bayar, Y.~Sarac and H.~Sundu,
Phys.\ Rev.\ D {\bf 80}, 096007 (2009).

P








\bibitem{Detmold:2015aaa}
W.~Detmold, C.~Lehner and S.~Meinel,
Phys.\ Rev.\ D {\bf 92}, no. 3, 034503 (2015).

\bibitem{Mannel:2015osa}
T.~Mannel and D.~van Dyk,
Phys.\ Lett.\ B {\bf 751}, 48 (2015).

\bibitem{Faustov:2016pal}
R.~N.~Faustov and V.~O.~Galkin,
Phys.\ Rev.\ D {\bf 94}, no. 7, 073008 (2016).
\bibitem{Gutsche:2015mxa}
T.~Gutsche, M.~A.~Ivanov, J.~G.~Körner, V.~E.~Lyubovitskij, P.~Santorelli and N.~Habyl,
Phys.\ Rev.\ D {\bf 91}, no. 7, 074001 (2015)
Erratum: [Phys.\ Rev.\ D {\bf 91}, no. 11, 119907 (2015)].

\bibitem{Dutta:2015ueb}
R.~Dutta,
Phys.\ Rev.\ D {\bf 93}, no. 5, 054003 (2016).


\bibitem{gikls}
T.~Gutsche, M.~A.~Ivanov, J.~G.~K\"orner, V.~E.~Lyubovitskij and P.~Santorelli,
Phys.\ Rev.\ D {\bf 90}, no. 11, 114033 (2014);
Phys.\ Rev.\ D {\bf 93}, no. 3, 034008 (2016).

\bibitem{prc}
M.~Pervin, W.~Roberts and S.~Capstick,
Phys.\ Rev.\ C {\bf 72}, 035201 (2005).

\bibitem{wkl}
H.~W.~Ke, X.~Q.~Li and Z.~T.~Wei,
Phys.\ Rev.\ D {\bf 77}, 014020 (2008).

\bibitem{Datta:2017aue}
A.~Datta, S.~Kamali, S.~Meinel and A.~Rashed,
JHEP {\bf 1708}, 131 (2017).



\bibitem{Dai:1996xv}
Y.~B.~Dai, C.~S.~Huang, M.~Q.~Huang and C.~Liu,
Phys.\ Lett.\ B {\bf 387}, 379 (1996).


\bibitem{Shifman:1978by}
M.~A.~Shifman, A.~I.~Vainshtein and V.~I.~Zakharov,
Nucl.\ Phys.\ B {\bf 147}, 448 (1979).


\bibitem{Reinders}
L.~J.~Reinders, H.~Rubinstein and S.~Yazaki,
Phys.\ Rept.\ \textbf{127}, 1 (1985).



\bibitem{Patrignani:2016xqp}
C. Patrignani et al. (Particle Data Group), Chin. Phys. C {\bf 40}, 100001 (2016) and 2017 update.

\bibitem{Dosch:1988vv}
H.~G.~Dosch, M.~Jamin and S.~Narison,
Phys.\ Lett.\ B {\bf 220}, 251 (1989).

\bibitem{Belyaev:1982cd}
V.~M.~Belyaev and B.~L.~Ioffe,
Sov.\ Phys.\ JETP {\bf 57}, 716 (1983) [Zh.\ Eksp.\ Teor.\ Fiz.\
{\bf 84}, 1236 (1983)].

\bibitem{Ioffe:2005ym}
B.~L.~Ioffe,
Prog.\ Part.\ Nucl.\ Phys.\  {\bf 56}, 232 (2006).


\bibitem{kk}
P.~Bialas, J.~G.~K\"orner, M.~Kramer and K.~Zalewski,
Z.\ Phys.\ C {\bf 57}, 115 (1993).



\bibitem{swd}
S.~Shivashankara, W.~Wu and A.~Datta,
Phys.\ Rev.\ D {\bf 91}, no. 11, 115003 (2015).



\end{thebibliography}
\end{document}